\documentclass[aip,apl,reprint]{revtex4-1}

\usepackage{bm}
\usepackage[reqno,intlimits]{amsmath}
\usepackage[latin1]{inputenc}

\begin{document}

\title{High Energy Gain in Three-Dimensional Simulations of Light Sail Acceleration}

\author{A.~Sgattoni}\email[ ]{andrea.sgattoni@polimi.it}
\affiliation{Dipartimento di Energia, Politecnico di Milano, Milano, Italy}
\affiliation{CNR, Istituto Nazionale di Ottica, u.o.s. ``Adriano Gozzini'', Pisa, Italy}
\author{S.~Sinigardi}
\affiliation{CNR, Istituto Nazionale di Ottica, u.o.s. ``Adriano Gozzini'', Pisa, Italy}
\affiliation{Dipartimento di Fisica e Astronomia, Universit\`a di Bologna, Bologna, Italy}
\affiliation{INFN sezione di Bologna, Bologna, Italy}
\author{A.~Macchi}
\affiliation{CNR, Istituto Nazionale di Ottica, u.o.s. ``Adriano Gozzini'', Pisa, Italy}
\affiliation{Dipartimento di Fisica ``Enrico Fermi'', Universit\`a di Pisa, Pisa, Italy}

\date{\today}

\begin{abstract}
The dynamics of radiation pressure acceleration in the relativistic light sail regime are analysed by means of large scale, three-dimensional (3D) particle-in-cell simulations. Differently to other mechanisms, the 3D dynamics leads to faster and higher energy gain than in 1D or 2D geometry. This effect is caused by the local decrease of the target density due to transverse expansion leading to a ``lighter sail''. However, the rarefaction of the target leads to an earlier transition to transparency limiting the energy gain. A transverse instability leads to a structured and inhomogeneous ion distribution.
\end{abstract}

%\keywords{}

\maketitle
The development of high power laser systems able to deliver short ultraintense-pulses drove an increasing interest to study laser-plasma interaction with particular focus on realizing compact sources of high energy electrons, ions and photons. In particular, several mechanisms of ion acceleration have been proposed and tested \cite{daidoRPP2012,*macchiRMP13} also thanks to progress in target manufacturing \cite{maNIMA11} and pulse contrast \cite{dromeyRSI04,*thauryNP07}. These latter allowed the first recent experimental investigations of radiation pressure acceleration (RPA) of thin solid foils, i.e. the so-called light sail (LS) regime \cite{henigPRL09b,*dollarPRL12,*kimPRL13,*aurandNJP13,*steinkePRSTAB13,palmerPRL12,karPRL12}. On the theoretical side, the LS configuration has been proposed and studied in the last ten years and it has been shown through simulations \cite{esirkepovPRL04} that LS becomes very efficient at intensities beyond $10^{23}~{\mbox{W cm}^{-2}}$ (foreseen with next generation facilities) in the regime where the ions become relativistic. 

In the basic one-dimensional (1D) picture of LS, the target, provided that its integrity and reflectivity are kept on a sufficiently long time scale, behaves almost as a perfect mirror and can be efficiently accelerated to relativistic velocities $V=\beta c$. The energy gain, though, after an early stage of exponential growth, becomes rather slow [$\gamma(t)\sim t^{1/3}$ where $\gamma=(1-\beta^2)^{-1/2}$] which would be an issue in a realistic three-dimensional (3D) scenario where the acceleration length required to obtain the maximum energy (for a given laser pulse) may exceed the diffraction length of the laser beam. However, it has been theoretically shown \cite{bulanovPRL10,*bulanovPoP10} that, in proper conditions and in a multidimensional case where the laser pulse has a finite focal spot, the decrease of the target areal density due to transverse expansion (so that the sail becomes effectively ``lighter'') may lead to a faster energy gain, i.e. $\gamma(t)\sim t^{3/5}$ in 3D geometry. A potentially ``unlimited'' energy gain is thus predicted, although at the expense of the number of accelerated ions. Possible limitations to the ``unlimited'' acceleration may come from the onset of target transparency (see Ref.  \citenum{PalaniyappanNPhys2012} and references therein) and from the development of a Rayleigh-Taylor instability \cite{palmerPRL12,pegoraroPRL07,*chenPoP11,*khudikPoP14,SgattoniArxiv14, *XuAPL2014, *ZhangAPL2012}.

The RPA-LS regime has been so far investigated extensively with particle-in-cell (PIC) simulations mostly in 1D and 2D (see e.g. Refs.\citenum{bulanovPRL10,*bulanovPoP10,klimoPRSTAB08,
*robinsonNJP08,*macchiNJP10,*yanPRL09,*qiaoPRL10,
*BadziakAPL2011,*BadziakAPL2012}), with few 3D studies having been performed \cite{esirkepovPRL04,tamburiniPRE12,yuPRL13,xuAPL14} mainly because of the very demanding computational requests. Since the scaling of ion energy with time, the diffraction length of the laser beam and the nonlinear RTI evolution are all dependent on the dimensionality, a comprehensive 3D investigation on long time scales is essential. A previous numerical work \cite{tamburiniPRE12} brought preliminary evidence of ion energies being higher in 3D than in lower dimensionality, but due to limited computing resources the simulations did not reach the end of the acceleration stage.

Here we present the results of large scale 3D simulations performed with the PIC code \texttt{ALaDyn} \cite{benedettiIEEE08}. We followed the LS dynamics in the ultrarelativistic regime until the end of the acceleration stage. We observed that in 3D the energy gain of the fastest ions is higher and faster than in 1D and 2D, and that the evolution of the maximum ion energy with time follows the power laws predicted by the analytical theory of Ref.\citenum{bulanovPRL10}. The ``unlimited'' acceleration is however limited by the onset of the target transparency. The shape of the accelerated ion bunches is not uniform but characterised by peculiar net-like structures attributed to the 3D dynamics of a transverse RTI \cite{SgattoniArxiv14}.

The 3D simulations of Ref.\citenum{tamburiniPRE12} indicated an ``operating point'', defined by the use of circular polarization (CP) of the laser pulse and by the target thickness $\ell_t$ matching the laser amplitude according to $\zeta=\pi(n_e/n_c)(\ell_t/\lambda) \simeq a_0$ where $a_0=(I/2m_ec^3n_c)^{1/2}$, $I$ and $\lambda$ are the laser intensity and wavelength, respectively, and $n_c=\pi m_ec^2/e^2\lambda^2$ is the cut-off density. In these conditions the light sail acceleration was shown to yield the highest ion energy while being relatively unaffected by both limited numerical resolution and inclusion of radiation friction effects. In order to extend the simulations on a much longer time scale we use a lower number of particles per cell after having verified that with such reduction the results of Ref.\citenum{tamburiniPRE12} are unaffected. In addition, we employ a non uniform grid in the transverse direction: a constant cell spacing is maintained in a region around the axis and then gradually stretched (using the tangent as stretching function) towards the edge. This allows us to keep a high resolution in the center and to  contain the expanding plasma with a reasonable number of grid points. The simulation box is $93\lambda$ wide along $x$ (the laser-propagation direction) and $120\lambda$ along $y$ and $z$. In the central region ($93\times 60\times60\,\lambda$) the cell size is $\Delta x=\lambda/44$, $\Delta y=\Delta z=\lambda/22$.  The grid size is $4096\times1792\times1792$ cells and 64 macro particles per cell per species are used accounting for a total number of about $2\times 10^{10}$. The simulations were run on 16384 BlueGene/Q cores on FERMI at CINECA (Bologna, Italy).
\begin{figure}[tb]
\centering
\includegraphics[width=0.49\textwidth]{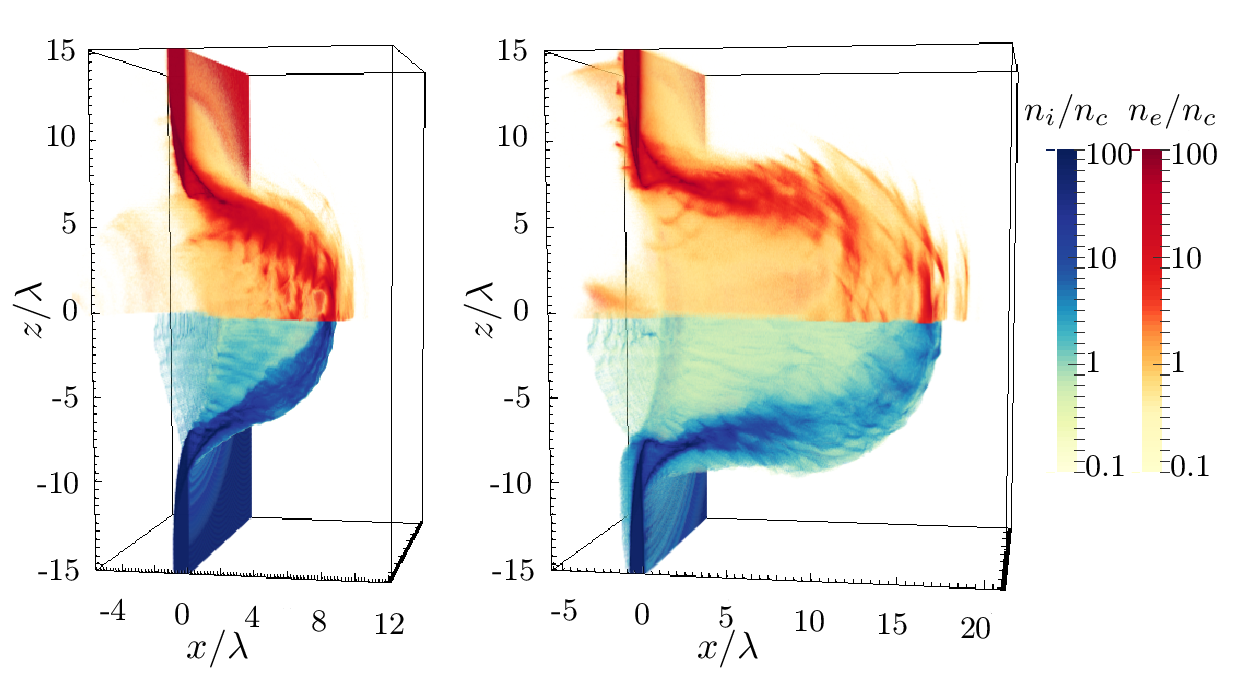}
\caption{Left ($t=20T$) and right ($t=30T$) panel: 3D snapshots of the density of electrons (red tones, upper half, $z>0$) and Carbon ions (blue tones, lower half, $z<0$). Only the $y>0$ region is shown for visualization purposes. . 
\label{fig_3D}}
\end{figure}

In the following we mostly focus on a target composed of a first layer of ions with charge to mass ratio $Z/A=1/2$ (e.g. C$^{6+}$), thickness $\ell_t=\lambda$ and initial electron density $n_e=64n_c$, and a second layer of protons, having thickness $\ell_r=\lambda/22$ and density $n_e=8n_c$. For reasons of computational feasibility, the density is lower than for real solid targets (for comparison Carbon targets have a mean electron density around $400n_c$) but the areal density has a realistic value (for Diamond-like Carbon foils the thickness may be down to $\simeq \lambda/100$). The target configuration mimics a Carbon foil with hydrogen contaminants on the rear side and allows to differentiate the dynamics of different charge states. The peak normalized amplitude of the laser field corresponding to the ``optimal'' thickness condition $a_0 \simeq \zeta$ is $a_0=198$. In all simulations, the laser pulse has a transverse Gaussian profile with waist diameter $w=6\lambda$ and a longitudinal $\cos^2$-like profile with a FWHM duration $\tau_p=9T$ (where $T=\lambda/c$ is the laser period), all referred to the profile of the fields. The simulations have been run for a time $t=80T$, where $t=0$ corresponds to the moment when the laser pulse front reaches the edge of the target. 

Fig.\ref{fig_3D} shows density snapshots for both electrons and ions at intermediate stages of the acceleration process. The side view shows the strong typical ``cocoon'' deformation of the target. The electron density shows structures with longitudinal modulation on the scale of $\lambda$, similar to those observed in Refs.\citenum{tamburiniPRE12,yuPRL13}.

\begin{figure}[bt]
\centering
\includegraphics[width=0.48\textwidth]{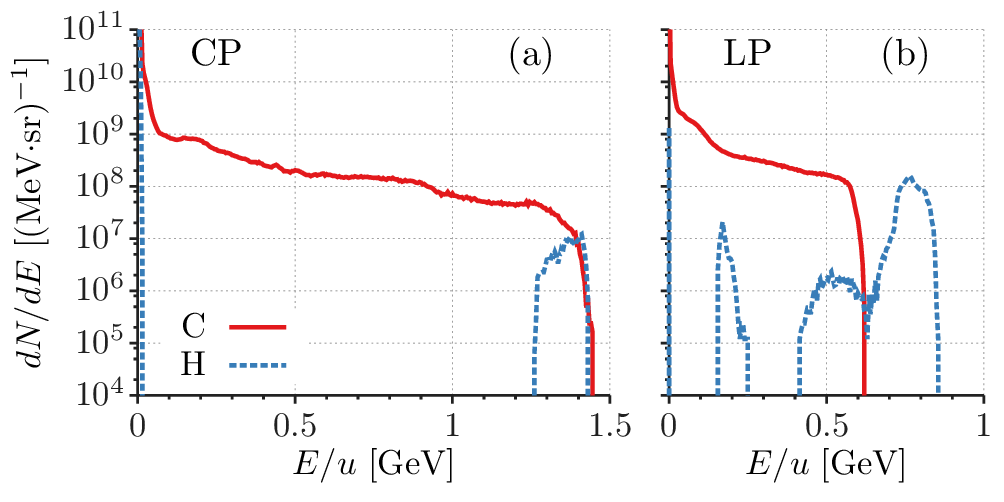}
\caption{Energy spectra for carbon ions (solid red) and protons (dashed blue) at the end ($t=80T$) of the simulations with circular (CP) and linear (LP) polarization, for ions within a cone with semi-aperture 5 of degrees around the $x$-axis. The absolute numbers in MeV$^{-1}$sr$^{-1}$ have been obtained assuming $\lambda=0.8\mu\text{m}$. 
\label{fig_spectra}}
\end{figure}

First we discuss the ion energy spectra at the end of the simulation of Fig.\ref{fig_3D}. As we are primarily interested in ions moving near the axis, we select particles whose momentum is within a cone with a semi aperture of 5 degrees. Inside this cone, the proton spectrum has a narrow peak, while the Carbon distribution is broad as shown in Fig.\ref{fig_spectra}~(a). The maximum energy per nucleon of carbon ions and protons are identical indicating that the most energetic particles of both species move at the same velocity. In contrast, for a 3D simulation identical to that of Fig.\ref{fig_3D} but with linear polarization (LP) instead of CP (and the peak field amplitude $a_0=198\sqrt{2}\simeq 280$ to keep the same mean intensity), the ions with different $Z/A$ ratio tend to form separate bands in the energy spectra, with the protons ending up with highest energy and the cut-off for heavier ions being in correspondence of the lower end of the proton peak, as shown in Fig.\ref{fig_spectra}~(b). This behavior is also observed in simulations at lower intensity ($a_0=100$, not shown), both for CP and LP.  Similar features were also found experimentally at much lower intensities \cite{karPRL12}. 
Assuming $\lambda=0.8\mu\text{m}$, we calculated the transverse rms emittance $\epsilon_{\perp}=(\sigma_{\perp}\sigma_{p_{\perp}}-m_{\perp}^2)^{1/2}$ (being $\sigma_{\perp},\,\sigma_{p_{\perp}},\,m_{\perp}$ respectively the variance of $r$ and of $p_{\perp}$ and the covariance of $(r,p_{\perp})$) of the ions having energy per nucleon $E>500\text{MeV/u}$ obtaining $4\times 10^{-2}$ and $8\times 10^{-2}\,\text{mm\,mrad}$ for carbon and hydrogen ions, respectively.
Such emittance values are compatible with experimental results on laser-driven acceleration at much lower intensities and proton energies \cite{cowanPRL04,*nuernbergRSI09} (a comparison with conventional accelerators is not straightforward due to the non-monochromatic nature of the ion bunches). 
\begin{figure}[t]%[htb]
\centering
\includegraphics[width=0.48\textwidth]{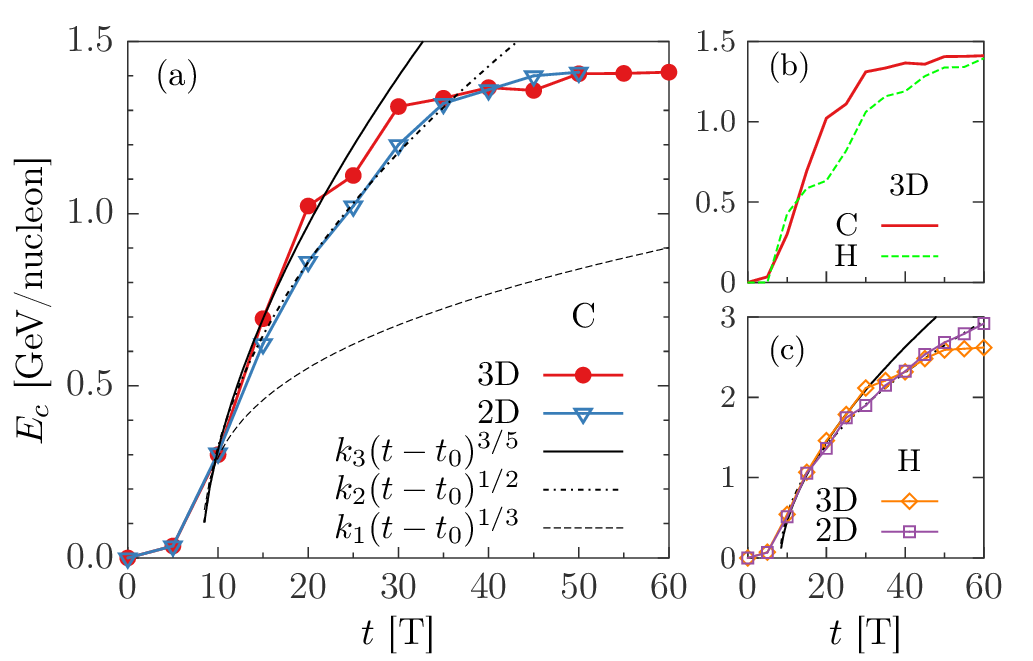}
\caption{(a): cut-off energy of C ions for the 3D simulation of Fig.\ref{fig_3D} (red circles) and a 2D simulation with same parameters (blue triangles). The continuous and dash-dotted line represent the fits with power law functions ($k_3=220$ and $k_2=245$). A similar function with the predicted 1D scaling is drawn for comparison with (dashed line). (b): same as (a), but for a pure hydrogen target ($k_3=330$ and $k_2=410$). (c) comparison between the energy of C ions and of protons in 3D for the same simulation of (a).
\label{fig_scaling}}
\end{figure}

We now consider the evolution of the cut-off energy of Carbon ions with time, shown in Fig.\ref{fig_scaling}~(a). Data for both the 3D simulation of Figs.\ref{fig_3D}-\ref{fig_spectra} and a 2D simulation with the same parameters are shown. In both cases, after an early stage of exponential growth ($t\lesssim 8T$) the time evolution shows an intermediate stage where the time dependence is well fitted by a power law (see below) until the acceleration saturates. Analysis of the simulations shows that such transition occurs when the target becomes transparent to the laser pulse. In the 3D case the energy rises with time faster than in 2D but the transition to transparency is also reached at an earlier time. Eventually the final energy values in 2D and 3D are very close, but both much higher than the 1D value. The cut-off energy of protons from the ``contaminant'' layer follows the temporal history of the energy of C ions closely. 

For the intermediate stage we used a power law $m_pc^2\gamma(t)=k_D(t-t_0)^{\alpha_D}$ as a fitting function for the energy with three free parameters ($k_D,\;t_0,\;\alpha_D$). The values of $\alpha_D$ are in very good agreement with the predictions of the analytical model of Ref.\citenum{bulanovPRL10} in the strongly relativistic, asymptotic limit (see also Ref.\citenum{macchiHPL14} for a simplified derivation):
\begin{equation}
\gamma(t)=\left(\frac{t}{\tau _D}\right)^{\alpha} \;,\quad \alpha=\frac{D}{D+2} \label{eq_energy}
\end{equation}
where $D=1,\,2,\,3$ is the dimensionality of the system and $\Omega=2I/\sigma_0c^2$. The time constants $\tau_D$ are
\begin{equation}
\tau_{1}=\left(\frac{3}{4\Omega}\right),\;
\tau_{2}=\left(\frac{1}{\Omega\overline{\omega}_0}\right)^{1/2},\;
\tau_{3}=\left(\frac{48}{125\Omega\overline{\omega}^2_0}\right)^{1/3}\label{eq_tau},
\end{equation}
where the parameter $\overline{\omega}_0$ is the related to the initial transverse momentum $p_{\perp}=m_pr(0)\overline{\omega}_0$, with $r(0)$ the initial position of the ions. In the model, after an initial ``kick'', the ions transverse motion is ballistic with constant momentum $p_{\perp}$, and the areal density of the target decreases due to the transverse expansion. 
An analytical estimate can be obtained considering the effect of the transverse ponderomotive force of a Gaussian laser pulse, $dp_{\perp}/dt\simeq -m_ec^2\partial_r\sqrt{1+\langle a^2(r,t)\rangle} $, where $\langle a^2(r,t)\rangle=a_0^2\exp(-2r^2/w^2)$. Near the focal axis ($r \ll w$) we have $dp_{\perp}/dt\simeq 2m_ec^2 a_0 r/w$, thus assuming an impulsive acceleration on a time $\Delta t$, we obtain $p_{\perp}\simeq m_e c^2({2r(0)}/{w^2})a_0 \Delta t$ and 
\begin{equation} 
\overline{\omega}_0 \simeq 2\frac{m_e}{m_p}\frac{a_0c^2\Delta t}{w^2} 
\; . \label{eq_omega0}
\end{equation}
\begin{figure}[tb]
\centering
\includegraphics[width=0.48\textwidth]{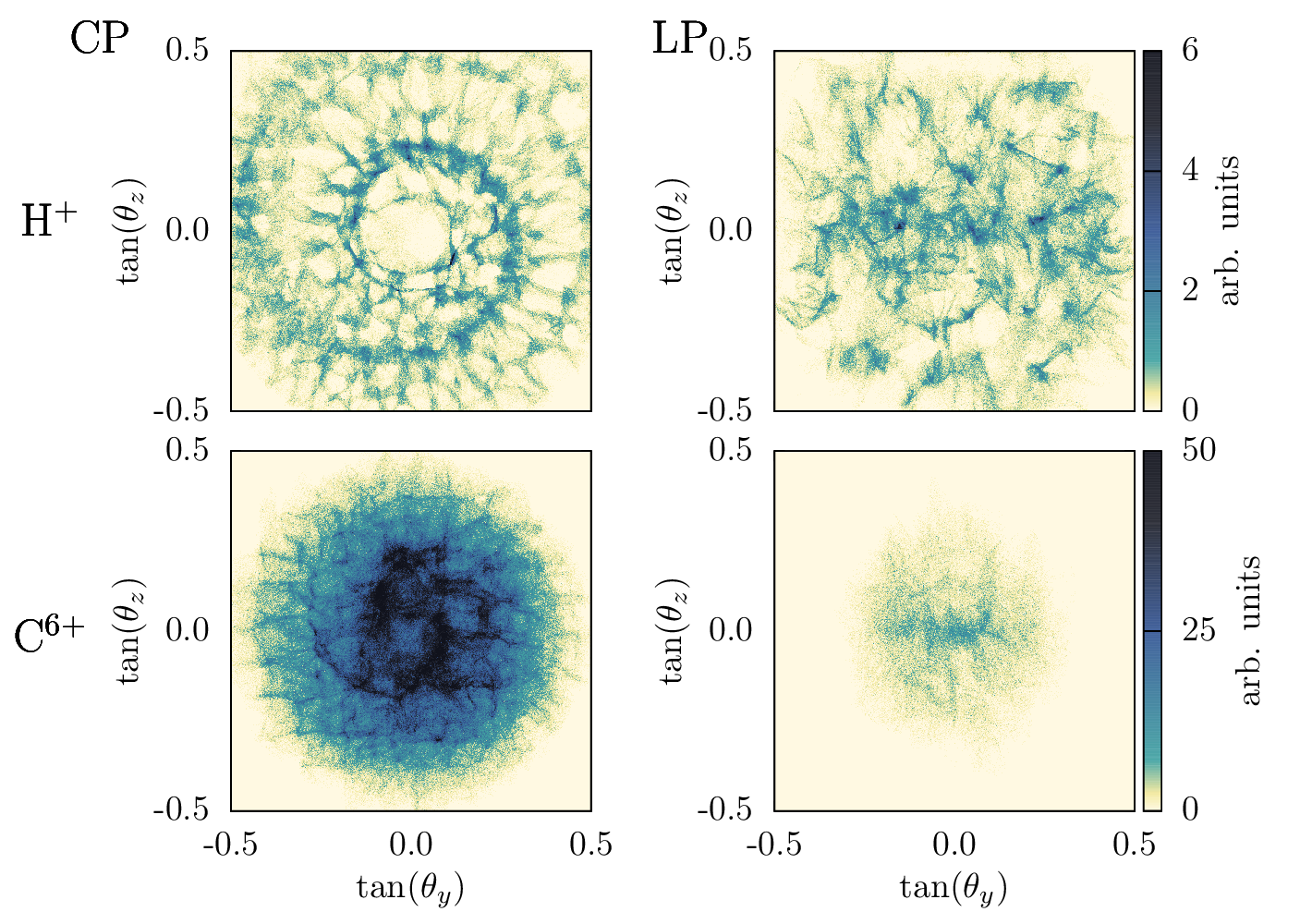}
\caption{Areal density of protons (P) and carbon (C) ions having energy $>500$MeV/u, obtained by projection of the $y,x$ plane at time $t=60$T, from the 3D simulation with circular polarization (CP) of Fig.\ref{fig_3D} and the corresponding simulation with linear polarization (LP). 
\label{fig_rcf}}
\end{figure}
The dependence of Eq.(\ref{eq_omega0}) from $\Delta t$ shows that a prompt acceleration favours the energy gain. Ref. \cite{bulanovPRL10} suggested the use of a properly shaped laser pulse to set an optimal condition for an "unlimited" acceleration which would be dependent on the dimensionality of the case considered and eventually increase the efficiency of the acceleration.
From the fit of Fig.\ref{fig_scaling} $\overline{\omega}_0\simeq 2.8\times 10^{-2}$ in the 3D case and  $\overline{\omega}_0\simeq 1.4\times 10^{-2}$ in 2D. The analytical estimate based on Eq.(\ref{eq_omega0}) gives $\overline{\omega}_0\simeq 6 \times 10^{-3}(\Delta t/T)$ which matches the simulation result if $\Delta t\simeq 4.7T$.
Simulations were also performed with a slab of hydrogen plasma $Z=A=1$ as in Refs.\citenum{esirkepovPRL04,tamburiniPRE12}, all other parameters being equal to Fig. \ref{fig_3D}. As shown in Fig.\ref{fig_scaling}~(c) the energy still follows closely the analytical scaling, reaching higher energies due to the lighter target; the values for the $\overline{\omega}_0$ parameters are $5.5\times 10^{-2}$ and $5.7\times 10^{-2}$ for the 3D and the 2D case, respectively. However, we notice that in the 2D case the scaling is followed for all the simulation time and the final energy is higher than for the 3D case. For the latter, the extra dimension allows a faster expansion which ``boosts'' the acceleration, but also leads to an earlier onset of transparency. Looking at the density distribution we observe that the transition to transparency occurs when the peak value is about $10n_c$.

Fig. \ref{fig_rcf} shows the density of particles obtained by projecting their trajectories on a plane. The image corresponds to what would be produced by the ion bunch at the plane of a detector such as a radiochromic film facing the rear side of the target.
A comparison between CP and LP for both $\text{C}^{6+}$ and $\text{H}^+$ shows how also for LP the signal density exhibits strong modulations but with a tendency of the structures to lengthen along the polarization direction. These structures are the result of the a RTI which develops already in the very early stages (not shown) in both the electron and ion densities. This analysis reveals the presence of structures in the beam, similarly to what was observed in an experiment of thin foil acceleration at lower intensity and longer pulse duration \cite{palmerPRL12}. These ``synthetic'' RCFs show how the polarization of the laser pulse affects the shape of the accelerated ions. The ion bunches are very laminar, as suggested by the low values of the emittance, but with a wide angular spread and non-uniform front.

In conclusion, three-dimensional simulations of light sail acceleration in the relativistic regime show that the energy gain versus time is much faster than in a one-dimensional model and show a scaling in very good agreement with the analytical theory of Refs.\citenum{bulanovPRL10}. The energy gain, however, is interrupted by the onset of target transparency which takes place earlier in 3D than in 2D, counterbalancing the advantage of the faster rarefaction. At the same time, the accelerated ion bunches, although very laminar, are strongly structured with a higly non-uniform front. These findings, obtained without considering specially shaped laser pulses or complex target geometries, are highly relevant to the design of next generation facilities for laser-driven ion acceleration at ultra-relativistic intensity.

We are grateful to F.~Pegoraro for enlightening discussions and to P.~Londrillo for help with the \texttt{ALaDyn} code. 
We acknowledge PRACE for access to the BlueGene/Q ``FERMI'', based in Italy at CINECA, via the project ``LSAIL''. Support from MIUR, Italy, via the FIR project ``SULDIS'' is also acknowledged.

%

%\bibliography{rpa-short}
\end{document}